\documentclass[trackchanges]{aastex63}

\graphicspath{{./}{figures/}}

\submitjournal{ApJL}

\shorttitle{Redshift Frame, Anisotropy, and Acceleration}
\shortauthors{Rubin et al.}

\begin{document}

\newcommand{\uhawaii}{\affiliation{Department of Physics and Astronomy, University of Hawai`i at M{\=a}noa, Honolulu, Hawai`i 96822}}

\author[0000-0001-5402-4647]{D. Rubin}
\uhawaii
\affiliation{E.O. Lawrence Berkeley National Laboratory, 1 Cyclotron Rd., Berkeley, CA, 94720}

\author{J. Heitlauf}
\uhawaii

\shortauthors{Rubin and Heitlauf} 
\title{Is the expansion of the universe accelerating? All signs {\it still} point to yes:
a local dipole anisotropy cannot explain dark energy}
 
\correspondingauthor{David Rubin}
\email{drubin@hawaii.edu}

\begin{abstract}

Type Ia supernovae (SNe~Ia) provided the first strong evidence that the expansion of the universe is accelerating. With SN samples now more than ten times larger than those used for the original discovery and joined by other cosmological probes, this discovery is on even firmer ground. Two recent, related studies (\citealt{nielsen16} and \citealt{colin19}, hereafter N16 and C19, respectively) have claimed to undermine the statistical significance of the SN~Ia constraints. \citet{rubinhayden16} (hereafter RH16) showed N16 made an incorrect assumption about the distributions of SN~Ia light-curve parameters, while C19 also fails to remove the impact of the motion of the solar system from the SN redshifts, interpreting the resulting errors as evidence of a dipole in the deceleration parameter. Building on RH16, we outline the errors C19 makes in their treatment of the data and inference on cosmological parameters. Reproducing the C19 analysis with our proposed fixes, we find that the dipole parameters have little effect on the inferred cosmological parameters. We thus affirm the conclusion of RH16: the evidence for acceleration is secure.

\end{abstract}


\section{Introduction} 
\label{sec:intro}

The discovery of the accelerated expansion of the universe \citep{riess98, perlmutter99} was originally made through the use of type Ia supernovae (SNe~Ia). With measurements of light-curve shape and color, the luminosity can be standardized so that the apparent magnitude can give the distance. Today, multiple cosmological probes combine together to provide a largely consistent picture of the universe on large scales \citep{planck18}.

A recent study \citep[][hereafter C19]{colin19} claims that this picture is wrong. They investigate the Joint Light-curve Analysis (JLA) SN Ia compilation \citep{Betoule_2014} and find evidence of a dipole anisotropy in the local deceleration parameter on $\sim 100$~Mpc scales. C19 claims that this dipole is larger and more statistically significant than the monopole acceleration, for which they find only weak evidence. C19 alleges that this dipole gives rise to falsely strong evidence of the accelerated expansion.

We reevaluate the C19 analysis and find four significant problems, each affecting their results or interpretation.

\begin{enumerate}
    \item C19's primary analysis makes the plainly incorrect assumption that SN Ia light-curve shape and color distributions are constant as a function of redshift (after selection cuts). That incorrect assumption comes from \citet{nielsen16} (hereafter N16), which C19 builds on. However, \citet{rubinhayden16} (hereafter RH16) has already demonstrated the inaccuracy of this assumption, discussing the need for redshift-dependent observed populations. We summarize that discussion, C19's new (since N16) arguments in favor of this incorrect assumption, and counter C19's objections using statistical significance and physically motivated reasoning. Incorrectly assuming redshift-independent observed distributions has little effect on inferring the dipole parameters, but causes large biases in the monopole cosmological parameters. The net result of these biases weakens a portion of the standardization of distant SNe, moving these SNe brighter and thus biasing the inferred cosmology towards less acceleration.

    \item Shockingly, C19 use {\it heliocentric} redshifts to compute their comoving model distances, leaving the well-established motion of the solar system with respect to the cosmic microwave background to imprint on the SN redshifts. They further use SNe as close as redshift 0.01, where the decision to use heliocentric or CMB redshifts affects the distance modulus at up to (5/log(10))(370 km/s)/(0.01 $c$) $\approx\pm 0.27$ magnitudes (see e.g., \citealt{davis11}). This distance modulus difference is dominant over per-SN distance uncertainty and is correlated across the sky. In addition to using heliocentric redshift, C19 removed the peculiar-velocity covariance terms from the JLA distance-modulus-uncertainty covariance matrix. This is in keeping with the spirit of using heliocentric redshifts, as C19 did not apply a peculiar-velocity model. However the removal of these covariance terms has a significant impact. When the full JLA peculiar-velocity covariance matrix is included in the analysis, we find no statistically significant anisotropy at $2 \sigma$, {\it even when using C19's preferred heliocentric redshift}. We show that these two related decisions are the primary driver of C19's claimed dipole anisotropy.
    
    C19's justification of the heliocentric choice is weak, citing evidence that bulk flows may exist on larger-than-expected scales. However, the choice to work in heliocentric redshifts is arbitrary; C19 shows no evidence that the heliocentric frame is closer to the average reference frame of nearby SN hosts than the CMB frame, let alone that it is better than employing the CMB frame plus corrections for known peculiar velocities (as the JLA analysis did). If the authors of C19 are concerned about the impact of peculiar velocities, then it makes very little sense to remove the peculiar-velocity-uncertainty covariances from the analysis.

    \item C19 calls out the hemispheric imbalance of surveys included in the JLA compilation; most SNe included are from Northern-hemisphere telescopes. However, C19 completely ignores consistent cosmological results from SNe published more recently than JLA that have been obtained with Southern-hemisphere telescopes (Carnegie Supernova Project \citealt{krisciunas17} and the Dark Energy Survey \citealt{desSN}). All of these data are public. C19 also ignores their own analysis's failure to find a strong correlation between the dipole term and the monopole term (their Figure~3), which suggests that choice of hemisphere does not affect the JLA analysis.
    
    \item The C19 preferred model (Equation~\ref{eq:dipole}, which they use for all their quoted results) includes a dipole anisotropy suppressed with an exponential falloff in redshift, and is thus 
    pathological when the scale of the exponential is poorly measured. As the supernova measurements are at finite redshift, extrapolating back to the redshift 0 dipole yields a strong degeneracy between the scale of the exponential ($S$) and the dipole value. We find that we must restrict the range of $S$ around the best-fit range found by C19 to achieve reasonable frequentist coverage (e.g., 68\% of the time, the true answer is in the 68\% credible interval).
    
\end{enumerate}

Our response is structured as follows. Section~\ref{sec:analysis} discusses our version of the C19 analysis, based heavily on the work of RH16. Section~\ref{sec:cosmo} shows our cosmological constraints. We run analyses with heliocentric redshifts, CMB-centric redshifts, and CMB-centric redshifts with the peculiar-velocity model JLA used \citep{hudson04, conley11} removed. We further investigate the impact of the peculiar-velocity covariances that C19 removed. Section~\ref{sec:conclusion} summarizes and concludes.

\section{Analysis and Reanalysis} \label{sec:analysis}

This section begins by reviewing kinematic parameters and outlining the C19 cosmological model (Section~\ref{sec:kinematic}). Section~\ref{sec:inference} describes our implementation of the analysis. Finally, Section~\ref{sec:population} discusses the importance of allowing the observed SN population distributions (after selection) to change with redshift, and dismantles the C19 claim that constant population distributions are preferred.

\subsection{Kinematic Parameters} \label{sec:kinematic}

The origin of the kinematic parameters is a series expansion of the scale factor $a(t)$ around the present time $t=t_0$:

\begin{equation} \label{eq:aoftseries}
    a(t) = a_0 \left[ 1 + H_0 (t - t_0) - \frac{1}{2!} q_0 H_0^2 (t - t_0)^2 + \frac{1}{3!} j_0 H_0^3 (t - t_0)^3 + \mathcal{O}([t - t_0]^4)
    \right] \;.
\end{equation}
$a_0$ is frequently defined to be 1. $H_0$ is the Hubble constant (or the Hubble parameter evaluated at $t=t_0$), given by 
\begin{equation}
    H_0 \equiv \frac{\dot{a}(t)}{a(t)} \vert_{t = t_0} \;.
\end{equation}
$q_0$ is the deceleration parameter and is given by
\begin{equation}
    q_0 \equiv -\frac{\ddot{a}(t) a(t)}{\dot{a}^2(t)} \vert_{t = t_0} \; ;
\end{equation}
the negative sign is a historical convention from when the expansion of the universe was assumed to be decelerating, making $\ddot{a}(t)$ negative. $j_0$ is the jerk parameter and is given by
\begin{equation}
    j_0 \equiv \frac{\dddot{a}(t) a^2(t)}{\dot{a}^3(t)} \vert_{t = t_0}\;.
\end{equation}

C19 follows \citet{visser04} in using the following series expansion for luminosity distance
\begin{equation} \label{eq:theirredshift}
    d_L (z) = \frac{c z}{H_0} \left[1 + \frac{1}{2}(1 - q_0) z - \frac{1}{6} (1 - q_0 - 3 q_0^2 + j_0 - \Omega_k) z^2 \right] \;.
\end{equation}
As $d_L$ = $(1 + z_{\mathrm{heliocentric}})$(comoving distance), we slightly modify Equation~\ref{eq:theirredshift} so that it scales as $(1 + z_{\mathrm{heliocentric}})$:

\begin{equation}
    d_L (z,\ z_{\mathrm{heliocentric}}) = \frac{1 + z_{\mathrm{heliocentric}}}{1 + z} \frac{c z}{H_0} \left[1 + \frac{1}{2}(1 - q_0) z - \frac{1}{6} (1 - q_0 - 3 q_0^2 + j_0 - \Omega_k) z^2 \right] \;,
\end{equation}
where $z$ nominally is CMB-centric with peculiar-velocity corrections, but we also show comparisons using heliocentric and CMB-centric redshifts. We form the distance modulus as
\begin{equation}
    \mu(z,\ z_{\mathrm{heliocentric}}) = 5 \log_{10} \left[ \frac{d_L(z,\ z_{\mathrm{heliocentric}})}{10\ \mathrm{parsec}} \right] \;.
\end{equation}
The C19 model separates $q_0$ into a monopole $q_{0m}$ and a dipole $q_{0d}$ which C19 points in the direction of the CMB dipole ($\vec{n}_{\mathrm{CMB}}$) in the their baseline analysis.
\begin{equation} \label{eq:dipole}
    q_0 (z_{\mathrm{SN}}, \vec{n}_{\mathrm{SN}}) = q_{0m} + q_{0d} (\vec{n}_{\mathrm{SN}} \cdot \vec{n}_{\mathrm{CMB}})\ e^{-z_{\mathrm{SN}}/S} 
\end{equation}
$S$ is the scale (in redshift) over which the dipole term falls off.

\subsection{Parameter Inference} \label{sec:inference}

As noted above, much of our analysis follows RH16, including implemeting the model in Stan \citep{carpenter17} with Pystan (\doi{10.5281/zenodo.598257}). We use 40 chains, each of which draws 1,000 samples after 1,000 discarded burn-in samples. We publish our code as an update to the original RH16 analysis.\footnote{\url{https://github.com/rubind/SimpleBayesJLA}} We (and C19) use a simple Bayesian hierarchical model for linear standardization of SNe Ia \citep[c.f.,][]{gull89, march11}. We treat all parameters in a Bayesian framework, while C19 uses frequentist statistics for the cosmological parameters (and other global parameters). The posteriors on $q_{0d}$, $q_{0m}$, and $j_0$ are relatively Gaussian, so this is not a significant difference.

As discussed in Section~\ref{sec:intro}, the model in Equation~\ref{eq:dipole} has pathologies. As $S$ is poorly constrained (which we show in Section~\ref{sec:cosmo}), and the SNe constrain any dipole at finite redshift, there is a large degeneracy between $q_{0d}$ (which is the dipole value at redshift 0) and $S$. For ease of comparison to C19, we use their model for our results. To ensure good frequentist coverage of our inference (especially on $q_{0d}$, which is better constrained than $S$), we use simulated-data testing. We use the actual coordinates and redshifts of the nearby SNe ($z < 0.15$) in JLA, and simulate data with $q_{0d} = 8$ (as in C19) and $q_{0d} = 0$. Running many simulations, we find that we need to restrict $S$ to give reasonable frequentist coverage for the inference. We restrict $S$ to the range 0.02--0.03, matching the range of $S$ values from C19 when $q_{0m}$, $q_{0d}$, and $S$ are simultaneously inferred.

\subsection{Population Distributions} \label{sec:population}

\begin{figure}
    \centering
    \includegraphics[width = 0.6\textwidth]{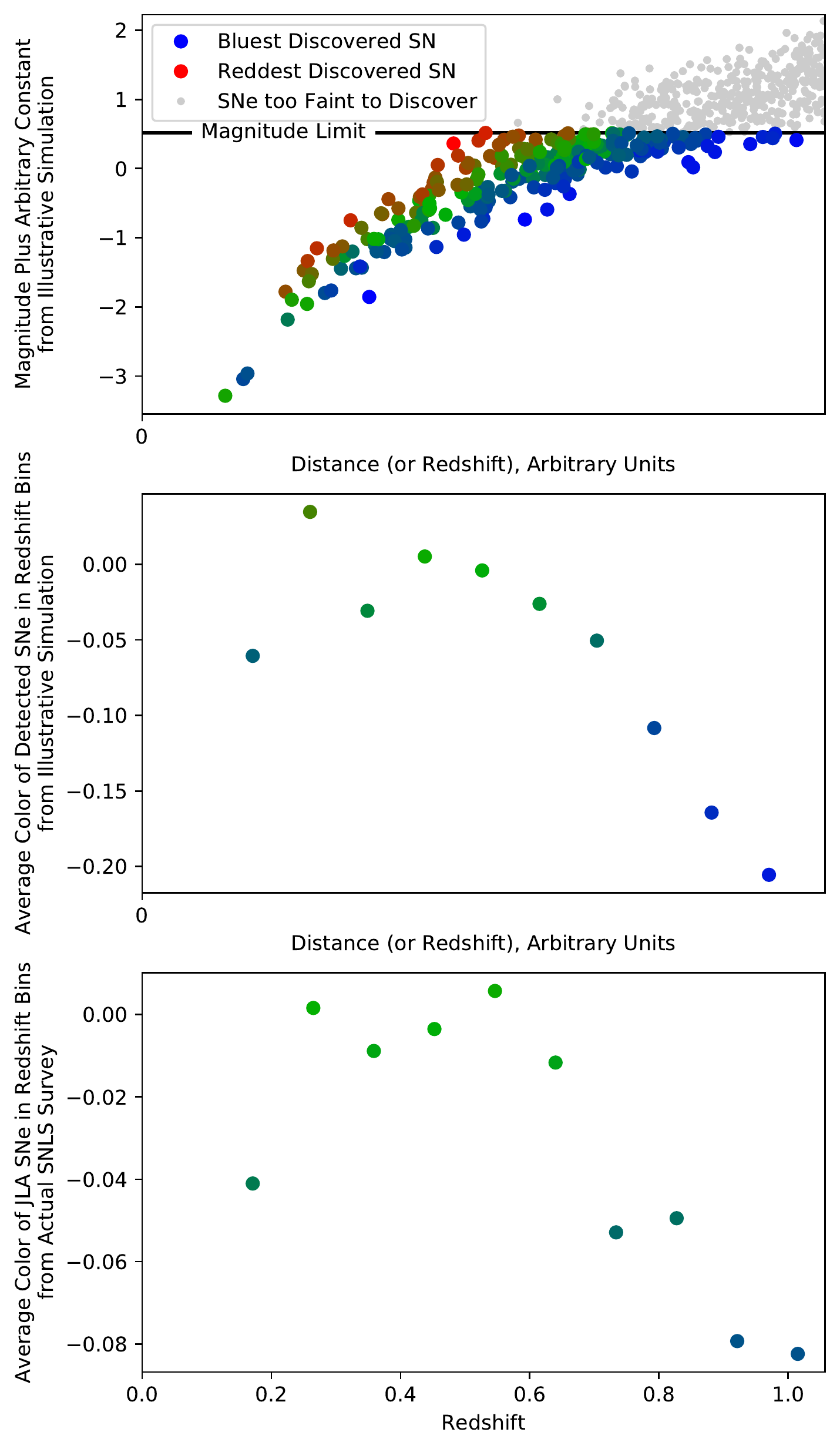}
    \caption{Illustration of the interaction of observed light-curve-parameter distributions and selection effects. We run a simple simulated survey containing only color and magnitude. This simulation assumes a distance modulus of $5 \log_{10}$(redshift) and a constant volumetric rate ($P(z) \propto z^2$), and is thus scale-invariant. In other words, both the redshift and distance are in arbitrary units, and the magnitudes are only defined up to an additive constant. The {\bf top panel} shows magnitude plotted against true distance (or redshift); the detected SNe (brighter than a cutoff magnitude) are plotted from blue to red according to their color. The SNe fainter than the cutoff magnitude are shown in light gray. This plot shows that magnitude correlates with both color and distance, and thus bluer SNe can be seen to much greater distances than redder SNe. The {\bf middle panel} shows average color of detected simulated SNe in bins. A clear trend towards detecting bluer SNe with greater distance is seen. The {\bf bottom panel} shows binned color vs. redshift for the real SNLS survey (see also RH16 Figure~1). The trend is qualitatively similar but quantitatively different (as a number of real-world effects were ignored in this simple simulation). This simple simulation shows that there is a strong expectation that the average color of detected SNe should vary significantly with redshift for magnitude-limited samples, even in the case where SNe have the same intrinsic population distribution at all redshifts. 
    \label{fig:colorvsz}}
\end{figure}

To illustrate the necessity of modeling the changes in the observed light-curve-parameter distributions with redshift, we simulate a very simple survey with just color and magnitude. The intrinsic color population is assumed to be Gaussian with width 0.1 magnitudes and mean 0 (constant in redshift). We generate redshifts with $P(z) \propto z^2$ (following a constant volumetric rate), then construct magnitudes using $5\log_{10}$(distance)~+~3.1~color~+~arbitrary~constant, assuming distance $\propto$ redshift. For illustrative purposes, we apply a hard magnitude limit, in contrast to actual selection probabilities, which are smooth with magnitude \citep[e.g.,][]{perrett10}. Figure~\ref{fig:colorvsz} shows the results of this simulation and qualitative agreement with the actual color trend with redshift in the SuperNova Legacy Survey in the JLA analysis. This tendency to select more luminous SNe out to larger redshifts is frequently referred to as a Malmquist bias \citep{malmquist}.

If one incorrectly assumes a redshift-independent population distribution (when the populations after selection do depend on redshift), this biases the latent values of light-curve shape and color for each SN towards the population mean \citep{woodvasey07, kessler09, karpenka2015supernova, rubin15b, rubinhayden16}. For color, this means closer SNe are biased bluer, and more distant SNe are biased redder. After applying the color standardization, closer SNe are biased fainter and more distant SNe are biased brighter. Such a bias thus undoes some of the evidence for the accelerated expansion.

RH16 introduced 12 additional parameters: for the nearby SNe, the Sloan Digital Sky Survey SNe, and the SuperNova Legacy Survey SNe, mean color and light-curve shape (after selection) are modeled as linear functions in redshift. Of course, the averages are not quite linear with redshift, but this is close enough over most of the redshift range of most of the SNe in each survey that more redshift-flexible models do not significantly change the cosmological inference (as noted in RH16).\footnote{An even better approach is to model both the selection process and the intrinsic population distribution as a function of SN sample and/or redshift. This can be done either as a simultaneous Bayesian model \citep{rubin15b} or with simulations \citep{scolnic16}.} C19 criticize the inclusion of these 12 additional parameters on three grounds, all of which are demonstrably false.

C19 claims that the additional fit parameters are a posteriori, in other words that these parameters were deliberately added by RH16 to the N16 analysis to recover the concordance cosmology. However, the importance of redshift-dependent color population distributions was well established much earlier than the N16 paper, e.g. in \citet{woodvasey07, kessler09, karpenka2015supernova, rubin15b}. The intrinsic light-curve shape population distribution is expected to be redshift-dependent as well, as there is a correlation between light-curve shape and host-galaxy properties \citep{hamuy95}.

C19 further claims that the additional parameters are not justified by the Bayesian information criterion \citep{schwarz1978}, a frequently used method that penalizes $-2$ log(likelihood) with $k \log(n)$ for $k$ parameters and $n$ observations. As discussed above, we have a strong expectation that the observed light-curve population distributions should be redshift dependent, negating this objection to the 12 additional parameters of RH16 even if these parameters were not favored by the Bayesian information criterion. The C19 implementation of the RH16 model gives a $-2$~log(likelihood) that ranges from $-298$ to $-332$, depending on other fit parameters. The C19 constant-populations-in-redshift model gives a $-2$~log(likelihood) that ranges from $-190$ to $-217$.\footnote{We validate these numbers from our model by computing the median and covariance of the average of the shape and color population distributions as a function of redshift, and then forcing those distribution means to be constant with redshift as C19 did. We find an effective $-2$~log(likelihood) preference for redshift dependence of 115, in line with the C19 values.} The Bayesian information criterion relative penalty for 740 measurements (as there are 740 SNe in the JLA compilation, so 740 light-curve-shape measurements for the six additional light-curve-shape parameters and 740 color measurements for the six additional color parameters) and 12 additional parameters is $12 \log(740) = 79$, much less than the $-2$~log(likelihood) difference. Even using $12 \log(740 \times 3) = 92$ ($\times 3$ for light-curve shape, color, and magnitude), still much less than the observed difference. We further note that a $\chi^2$ improvement of 115 for 12 degrees of freedom is a $\sim 9 \sigma$ improvement, again showing strong evidence for the RH16 model. Thus C19's own values show that the RH16 model is actually preferred {\it using the information criterion that C19 claims to be using}.

Finally, C19 criticizes RH16 for focusing on the light-curve shape and color parameters to the exclusion of magnitude. This claim is also false, as described in RH16. As part of the JLA analysis, \citet{Betoule_2014} included simulations of the impact of selection effects on each SN sample, removed its impact on magnitude, and propagated the uncertainties into the covariance matrix (much of this work was built on \citealt{conley11}). This bias is different from the deficient standardization that one gets from ignoring the redshift dependence of the population distributions, which is due to biasing the inferred latent values of light-curve shape and color. Intrinsic magnitude differences with redshift are controlled for by matching SNe in bins of host-galaxy stellar mass (as an easier-to-measure proxy for SN progenitor age or metallicity, e.g., \citealt{sullivan10}). In any case, the systematic uncertainties due to such effects are a small fraction of the signature of acceleration in the SN Ia data, and a fraction of the bias caused by ignoring the redshift dependence of the population distributions.

\section{Cosmological Constraints} \label{sec:cosmo}

\begin{figure}
    \centering
    \includegraphics[width = 0.9 \textwidth]{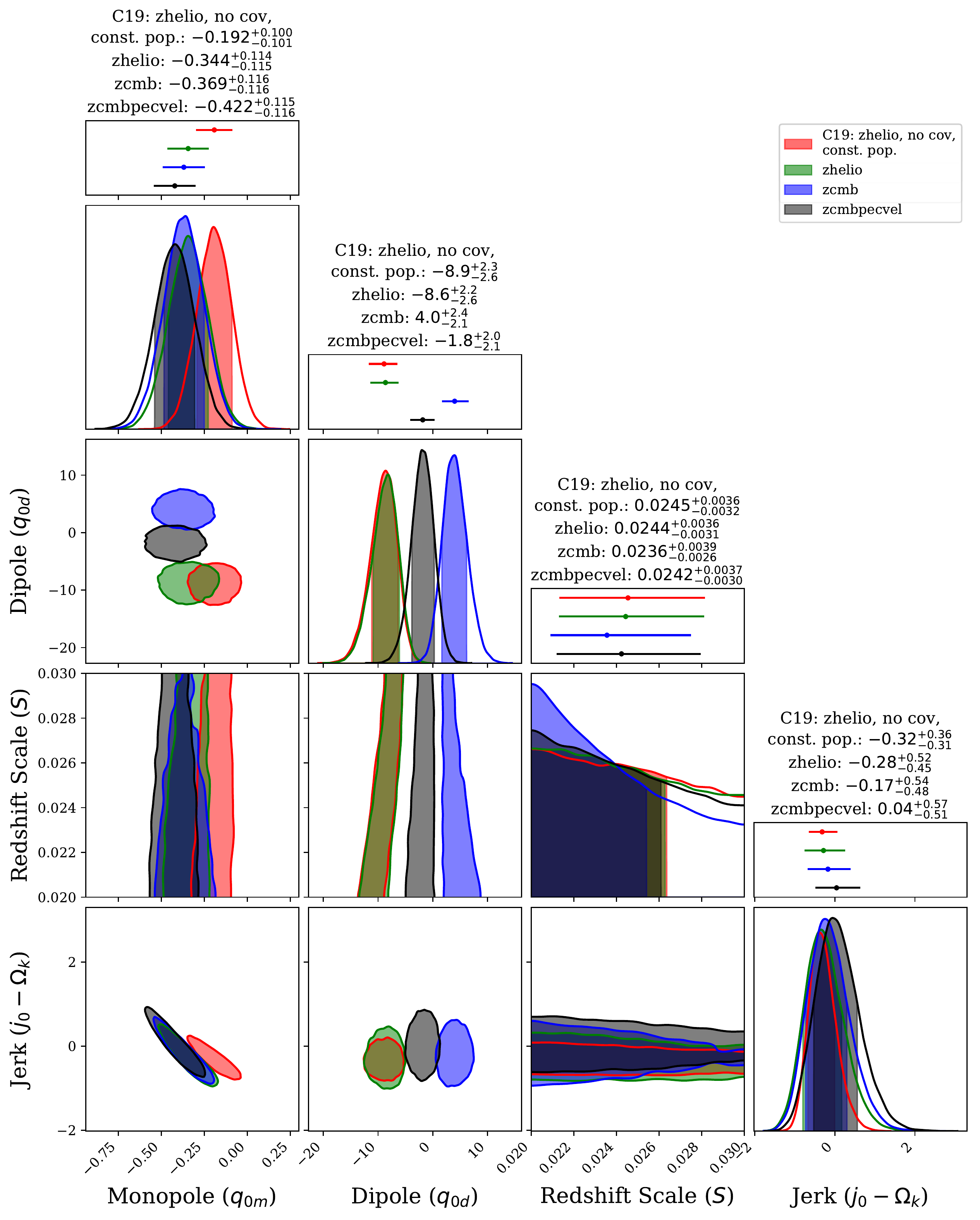}
    \caption{68.3\% credible regions and intervals for cosmological parameters. We show the results from the C19 assumptions (heliocentric redshifts, no peculiar-velocity-uncertainty covariances, and redshift-independent observed light-curve population distributions) in red. For our assumptions, we use all three considered redshifts: heliocentric (green), CMB-centric (blue), and CMB-centric with peculiar-velocity corrections (black), all of which have the RH16 redshift-dependent population model. We remove the peculiar-velocity covariances from the JLA covariance matrix to match C19. The cosmological parameter impacted the most by the redshift variants is $q_{0d}$. In particular, we see moderate evidence for a non-zero $q_{0d}$ dipole when heliocentric redshifts are used, but this evidence drops to $< 2\sigma$ for the other redshifts. The results using C19 assumptions are significantly offset due to their population model.
        \label{fig:cosmonocov}}
\end{figure}

\begin{figure}
    \centering
    \includegraphics[width = 0.9 \textwidth]{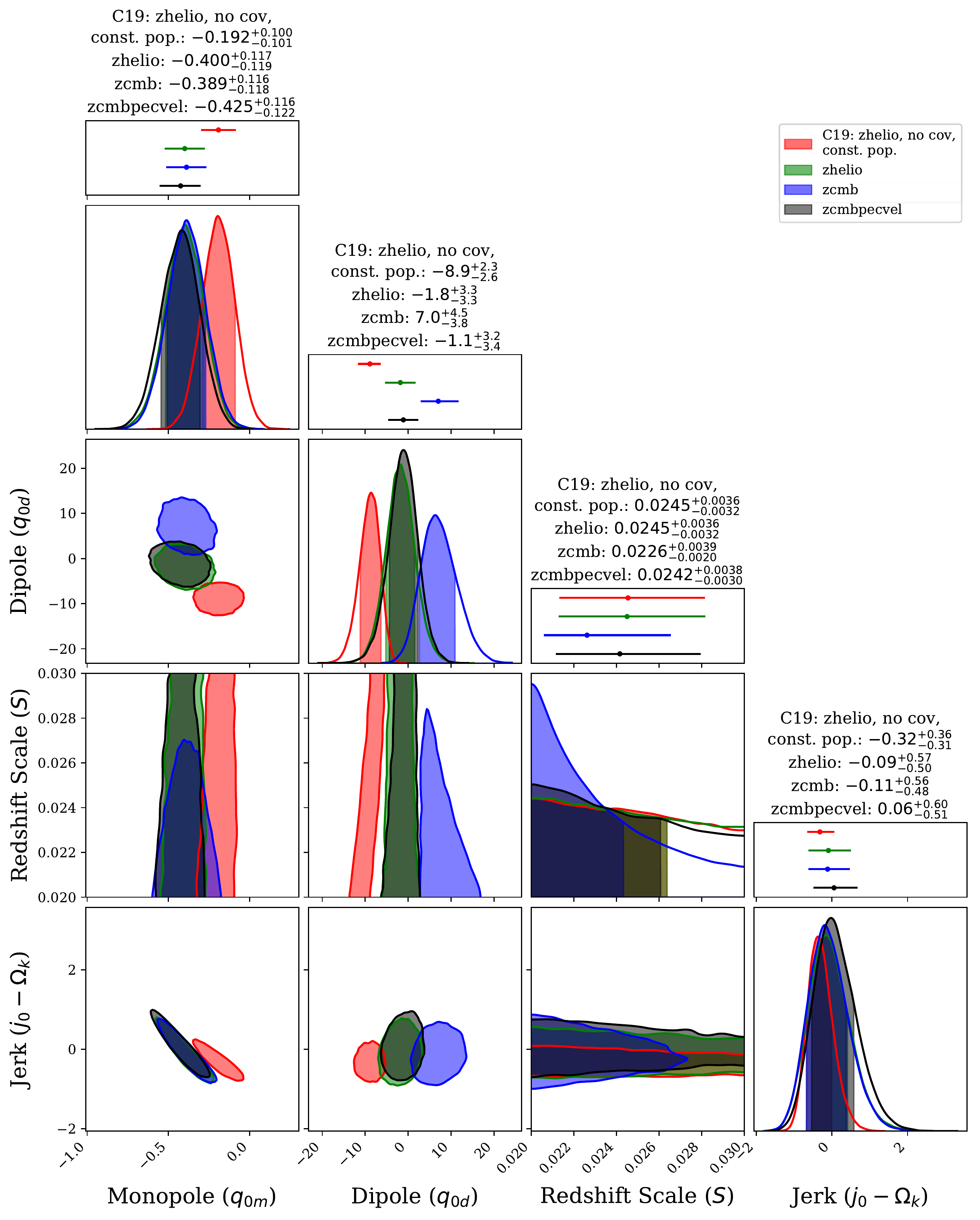}
    \caption{68.3\% credible regions and intervals for cosmological parameters. We show the results from the C19 assumptions (heliocentric redshifts, no peculiar-velocity-uncertainty covariances, and redshift-independent observed light-curve population distributions) in red. For our assumptions, we use all three considered redshifts: heliocentric (green), CMB-centric (blue), and CMB-centric with peculiar-velocity corrections (black), all of which have the RH16 redshift-dependent population model. Unlike Figure~\ref{fig:cosmonocov}, the peculiar-velocity-uncertainty covariances are included, as they were in the JLA analysis. In contrast to the results without the peculiar-velocity covariances, all of our cosmological results are consistent, with only a modest difference in $q_{0d}$. The results using C19 assumptions are again significantly offset.
        \label{fig:cosmowithcov}}
\end{figure}

Figures~\ref{fig:cosmonocov} and \ref{fig:cosmowithcov} show the 68.3\% credible regions and intervals for the cosmological parameters. We compute the credible regions by interpolating the posterior samples using kernel density estimation (KDE) and setting the contour level to enclose 68.3\% of the interpolated posterior.\footnote{The code is available at \url{https://github.com/rubind/kde_corner}.} Above the credible regions, we show the 1D PDFs. Here, we perform the analogous procedure in 1D: use a KDE (now 1D instead of 2D), and compute and fill the minimum distance that encloses 68.3\% of the PDF. Above the KDE results, we compute and show more conventional 1D constraints for each parameter that use the median, the 15.9th percentile, and the 84.1th. The KDE-based credible intervals will agree well with the percentile-based intervals for nearly Gaussian distributions. In contrast, these intervals can be seen to diverge for $S$, which shows a decreasing distribution.

Figure~\ref{fig:cosmonocov} shows constraints for three cases: heliocentric redshifts, CMB-centric redshifts, and CMB-centric redshifts with peculiar-velocity corrections. To match C19, we remove the peculiar-velocity covariances from the JLA covariance matrix. We see that the choice of redshift frame mostly affects $q_{0d}$. Figure~\ref{fig:cosmowithcov} shows the same cosmological constraints with the JLA peculiar-velocity covariance matrix included. In contrast to the Figure~\ref{fig:cosmonocov} results (without this part of the covariance matrix), $q_{0d}$ is consistent with zero at $2 \sigma$ for all three types of redshifts.

The N16/C19 assumption of redshift-independent observed light-curve population distributions underestimates cosmological uncertainties as well as shifting central values. This is primarily due to the increase in average uncertainty on SN color with redshift, from $\sim 0.02$ magnitudes at low-redshift to $\sim 0.06$ magnitudes at higher redshift. Both our analysis and C19 find that the observed color population distribution has an intrinsic width of 0.07 magnitudes, so when the mean of the color population is allowed to be redshift-dependent, this mean has significant uncertainty. When the color population distribution is incorrectly modeled as redshift-independent, the uncertainty on the mean is small (as the distribution is trained on many well-measured low-redshift SNe), resulting in smaller uncertainties on the high-redshift distance moduli. The impact is most evident in $j_0$, which is effectively inferred over a longer redshift baseline than $q_{0m}$. Figures~\ref{fig:cosmonocov} and \ref{fig:cosmowithcov} show $\sim$ 50--60\% larger uncertainties on $j_0$ when correctly using redshift-dependent population distributions.

It is worth briefly discussing the interpretation of series expansions (kinematic expansions) of the expansion history (Equation~\ref{eq:aoftseries}) to understand the strong correlation between $q_0$ and $j_0 - \Omega_k$. Frequently, we speak imprecisely about measuring $H_0$, $q_0$, or $j_0$ from SNe Ia (possibly combined with other probes). However, SNe measure distances at finite redshift and thus effectively measure these quantities at nonzero (but low) redshift. For example, the (statistically disfavored) not-currently-accelerating ($q_0$ forced to 0) models of C19 (in Table~A.1) have $j_0 - \Omega_k \sim -1.35$, and so would have experienced several Gyrs of acceleration in the recent past ($j_0 < 0$), even though the acceleration goes to zero today. See RH16 for a discussion of $q_0$ and $j_0 - \Omega_k$ JLA constraints using other (physical) models. 

\section{Conclusion} \label{sec:conclusion}

This work reimplements the C19 cosmological analysis to investigate their claims of a dipole term in the deceleration parameter ($q_{0d}$) and a statistically weak monopole ($q_{0m}$).

We show that the weak monopole finding is identical to the finding from the related N16 analysis, which was rebutted by RH16 for their incorrect use of constant-in-redshift SN populations (after selection). We find the same criticism still applies, and counter the C19 arguments against RH16, including finding an apparent miscalculation in the C19 Bayesian information criterion.

We find that the C19 result of a significant ($3.9 \sigma$) $q_{0d}$ depends strongly on both failing to remove our motion from the redshifts of SNe (that is, working in the heliocentric frame) and failing to include the JLA peculiar-velocity covariances. Changing either of these brings the evidence for a dipole below $2 \sigma$. Despite the inclusion of the dipole term, we see virtually the same constraints on $q_{0m}$ as RH16 saw on $q_0$ (this work: $q_{0m} = -0.425^{+0.116}_{-0.122}$, RH16: $q_0 = -0.425^{+0.115}_{-0.117}$) when using the same type of redshift model and using the same covariance matrix. We thus conclude that concerns over the value of $q_{0d}$ have little effect on the strength of the evidence for acceleration.

\acknowledgments

The technical support and advanced computing resources from the University of Hawaii Information Technology Services – Cyberinfrastructure are gratefully acknowledged. We thank Greg Aldering, Benjamin Rose, Brian Hayden, and the anonymous referee for careful feedback.

\software{
Matplotlib \citep{matplotlib}, 
Numpy \citep{numpy}, 
pystan (\doi{10.5281/zenodo.598257}),
Python,
SciPy \citep{scipy}, 
Stan \citep{carpenter17}}

\end{document}